\documentclass[aps,pra,twocolumn,superscriptaddress,superscriptaddress]{revtex4}

\usepackage{braket}
\usepackage{amsmath}
\usepackage{amsfonts}       
\usepackage{graphicx}
\usepackage{dcolumn}
\usepackage{epsfig}
\usepackage{bm}
\usepackage{tabularx}
\usepackage{array}
\usepackage{float}
\usepackage{longtable}
\usepackage{hyperref}
\usepackage{float}
\hypersetup{
colorlinks=false,
citecolor=blue,
linkcolor=blue,
urlcolor=blue,
}

\usepackage{amssymb}
\usepackage{bbm}
\usepackage[english]{babel}
\usepackage{setspace}
\usepackage{wasysym}
\usepackage{adjustbox}
\usepackage{framed}

\usepackage{latexsym} 
\usepackage{multirow}    
\usepackage{booktabs}    
\usepackage{calc}        
\usepackage[normalem]{ulem}

\usepackage{calrsfs}								
\DeclareMathAlphabet{\pazocal}{OMS}{zplm}{m}{n}	

\newcommand{\be}{\begin{equation}}
\newcommand{\ee}{\end{equation}}

\newcommand{\beq}{\begin{eqnarray}}
\newcommand{\eeq}{\end{eqnarray}}

\begin{document}

\title{Observation of the Quantum Zeno Effect on a NISQ Device}

\author{Andrea Alessandrini}
\affiliation{Dipartimento di Scienze Fisiche e Chimiche, Universit\`a dell'Aquila, via Vetoio,
I-67100 Coppito-L'Aquila, Italy.}

\author{Carola Ciaramelletti}
\affiliation{Dipartimento di Ingegneria e Scienze dell'Informazione e Matematica, Universit\`a dell'Aquila, via Vetoio,
I-67100 Coppito-L'Aquila, Italy.}
\affiliation{Dipartimento di Scienze Fisiche e Chimiche, Universit\`a dell'Aquila, via Vetoio,
I-67100 Coppito-L'Aquila, Italy.}

\author{Simone Paganelli}
\email[correspondence at: ]{simone.paganelli@univaq.it}
\affiliation{Dipartimento di Scienze Fisiche e Chimiche, Universit\`a dell'Aquila, via Vetoio,
I-67100 Coppito-L'Aquila, Italy.}

\begin{abstract}
We study the Quantum Zeno Effect (QZE) on a single qubit on IBM Quantum Experience devices under the effect of multiple measurements.  
We consider two possible cases: the Rabi evolution and the free decay. SPAM error mitigations have also been applied. In both cases we observe the occurrence of the QZE as an increasing  of the survival probability with the number of measurements. 
\end{abstract}

\maketitle

\section{Introduction}
\label{introduction}

The QZE refers  to the freezing of a system in a given state, or in a given subspace of the Hilbert space,  due to frequent measurements applied during its evolution. 
The possibility that the lifetime of an unstable system could depend on the measuring process was first analyzed in \cite{degasperis1974} and then subsequently formalized in \cite{misra1977,chiu1977}, where the QZE name was given to this phenomenon.
Since then, the QZE has attracted a lot of interest  \cite{ joos1984,cook1988,nakazato1996, presilla1996,facchi2009, pascazio2014, giacosa2017, lucia2023} and it has been observed experimentally in different contexts   \cite{itano1990, kwiat1995, nagels1997, drewsen2000, fischer2001, zhu2014}.

The QZE can be also due by the interaction with an external environment and produce a confinement  into a  subspace  \cite{facchi2002}. In this case, the coupling induces a dynamical superselection rule splitting the Hilbert space  of the open system into non-communicating quantum Zeno subspaces where coherent evolution can take place. Therefore, the study of the QZE is of great interest in quantum computation since it can be employed to control decoherence \cite{ facchi2004, facchi2006, facchi2009, facchi2010,martin2018, facchi2005}, and  to protect specific quantum states from decoherence by using projective measurements \cite{paz2012zeno}. 

Given an open system in an excited state, it starts to decay exponentially after a characteristic time $T_E$.  
For  times shorter than $T_E$, quantum effects prevail, the evolution deviates from the exponential behavior \cite{ chiu1977, pascazio2014, giacosa2017, fonda1978, pascazio1996} exhibiting a quadratic behavior: here QZE can occur.  Nevertheless, the coupling with a continuous spectrum could give rise to extra contributions which must be added to the quadratic one that could possibly produce a speed up of the decay rate. This phenomenon is the so-called Quantum Anti-Zeno Effect (QAZE) \cite{antoniou2001, kofman2001,facchi2001,na2021, lewenstein2000}.  Both the QZE and QAZE are characterized by two well defined times $T_Z$ and $T_{AZ}$, respectively, where generically $T_E>T_{AZ}>T_Z$. If the time between measurements is much shorter than $T_Z$, then the QZE can be observed, while in the time region between $T_Z$ and $T_{AZ}$ the QAZE exists \cite{antoniou2001}.

Protection of quantum states or subspaces of open systems from decoherence is essential for robust quantum information processing and quantum control.
In  \cite{paz2012zeno}, it was proven that the Weak-Measurement Quantum Zeno Effect (WMQZE) can protect appropriately encoded arbitrary states to arbitrary accuracy, while at the same time allowing for universal quantum computation or quantum control.

Quantum Zeno Effect can also be employed to suppress the failure events that would otherwise occur in a linear optics approach to quantum computing. From a practical viewpoint, that would allow the implementation of deterministic logic gates without the need of ancilla photons or high-efficiency detectors. Furthermore, QZE can lead  photons to behave as if they were fermions instead of bosons, which leads to a new paradigm for quantum computation \cite{franson2004quantum}. 

The Zeno effect has also been proposed as a method of quantum error correction \cite{zurek1984reversibility, zbarenco1997stabilization, vaidman1996error} or error reduction \cite{pachos2002quantum, ralph2003quantum}, and it is similar to the "bang-bang" method of error reduction \cite{viola1998, vitali1999, viola1999, protopopescu2003}.

Concerning the possibility to observe the QZE on a Noisy Intermediate Scale Quantum (NISQ)  device \cite{bharti2022}, simulations  have been performed with ideal (noiseless and decoherence free) qubits \cite{barik2020, thesis}, and predictions have been formulated about the behavior of real ones \cite{dominik2018, matsuzaki2010}. Furthermore, there are experiments involving superconducting flux qubits \cite{ kakuyanagi2015} and about the observation of the QZE from measurement controlled qubit-bath interactions involving a transmon circuit \cite{harrington2017}. 

In this paper, we show the observation of the QZE  on real NISQ devices. In particular,  we employed  IBM Quantum Experience devices, which belong to the class of gate based quantum computers, consisting in superconducting  Josephson junctions in  transmonic regime \cite{ devoret2004, kockum2019, koch2007}. In Section \ref{evolution} we describe the general problem of the QZE in two cases: the Rabi evolution and the free decay of a single qubit and how to ideally implement them on  the online platform IBM Quantum Experience. In Section \ref{results}, we give more insights about the implementations on real devices. Despite the big amount of noise, our results show a manifestation of the QZE in both the dynamics, with qualitative agreement with the theory.  In Section \ref{ch:mitigation} we  address the problem of  the noise, providing mitigations to the errors occurring in the  preparation of the initial state and in the measurement of the final state (SPAM errors). Moreover, additional data about those mitigations are shown in App. \ref{extrarun} and App. \ref{osaka}. Eventually, Section \ref{ch:conclusion} is devoted to our conclusions.

\section{Rabi evolution and free evolution}
\label{evolution}

\begin{figure*}
\includegraphics[width=0.8\textwidth]{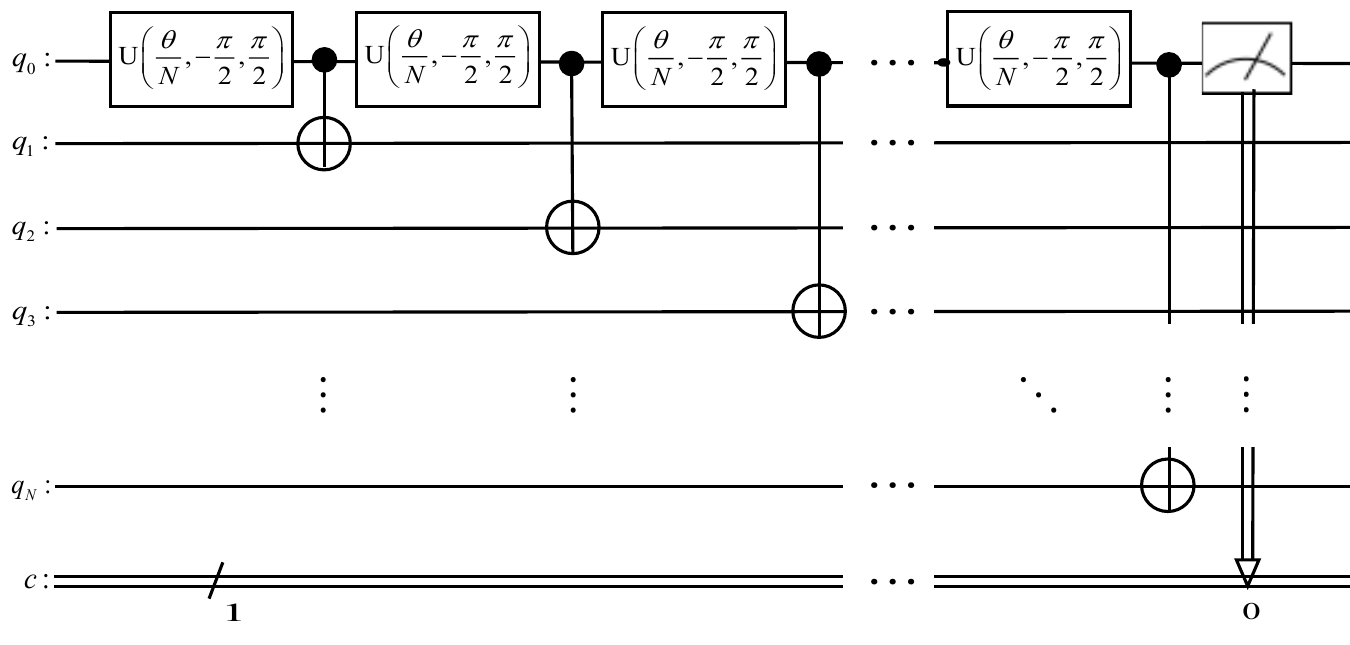}
\caption{Schematic circuit employed in  the QASM simulations. To simulate Rabi oscillations, we employ the rotation in Eq. \eqref{u} where  $\lambda=-\phi=\frac{\pi}{2}$ is fixed and $\theta$ plays the role of the evolution time. For the  $N$ intermediate measurements during the evolution  we need the same number of ancillary qubits. So the full rotation is divided into $N$ equivalent rotations, and each one is followed by a $CNOT$ gate.}
\label{scheme}
\end{figure*}

In the present section, we describe  two different time evolutions of a single qubit that we consider in this paper: the oscillations under a Rabi Hamiltonian and the free decay.  The first case was already studied  in \cite{barik2020} and \cite{thesis}, where  the Quantum Zeno dynamics has been found by IBM Quantum Experience simulations.  Here we briefly recall the model Hamiltonian and the measurement scheme, and in the next section we will compare the results of these simulations with the data obtained from a real IBM quantum computer.
The  Rabi Hamiltonian  for a single qubit is
\begin{equation}
\hat{H}=\hbar \omega\left(\ket{0}\bra{1}+\ket{1}\bra{0}\right)=\hbar \omega\hat{\sigma}_x,
\end{equation}
corresponding to  the time evolution operator 
\begin{equation}
\hat{U}(t,0)=e^{-\frac{i}{\hbar}\hat{H}t}=e^{-i\omega t\hat{\sigma}_x}=\cos(\omega t)\hat{\mathbb{I}}-i\sin(\omega t)\hat{\sigma}_x.
\end{equation}
Initializing the state of the qubit on $\ket{0}$, it evolves into an oscillation between the two logic states with frequency $\omega$
\begin{equation}
\ket{\psi\left(t\right)}=\hat{U}(t,0)\ket{0}=\cos(\omega t)\ket{0}-i\sin(\omega t)\ket{1}.
\end{equation}
The probability for the qubit to be found on its initial state (survival probability) will be denoted by 
\begin{equation}
p\left(t\right)=\left| \braket{0|\psi\left(t\right)}\right|^2.
\end{equation}
For $t\ll\frac{1}{\omega}$,  if  $N$ projective measurements are performed at intervals $\Delta t=\frac{t}{N}$, the time evolution changes into
\begin{equation}
p_{QZE}(t)=\left[p\left(\frac{t}{N}\right)\right]^N\simeq \left(1-\omega^2\frac{t^2}{N^2}\right)^N\xrightarrow{N\gg 1} e^{-\frac{\omega^2 t^2}{N}},
\label{e4}
\end{equation}
leading  to the QZE. 
The oscillations driven by the Rabi evolution have been implemented using the single-qubit rotation gate 
\begin{equation} 
\hat{U}_{\theta,\phi,\lambda}=\left[
                    \begin{matrix}
                        \cos\frac{\theta}{2} & -e^{i\lambda}\sin\frac{\theta}{2} \\
                        e^{i\phi}\sin\frac{\theta}{2} & e^{i(\phi+\lambda)}\cos\frac{\theta}{2}
                    \end{matrix}
                    \right],
\label{u}
\end{equation}
and fixing $\phi=-\frac{\pi}{2}$, $\lambda=\frac{\pi}{2}$ and $\theta=2\omega t$. To emulate the measurement we did not use the corresponding irreversible gate, but we employed  $CNOT$ gates and the entanglement among the states of different qubits. Given a normalized state for  the system qubit $\ket{\psi}=\alpha\ket{0}+\beta\ket{1}$,  its density matrix is 
\begin{equation} 
  \hat{\rho}_{\text{in}}=\ket{\psi}\bra{\psi}=\left[\begin{matrix}
            |\alpha|^2 & \alpha \beta^* \\
            \alpha^*\beta & |\beta|^2
        \end{matrix}\right].
\end{equation} 
Introducing an ancillary target qubit in the state $\ket{0}$, the action of the  $CNOT$ gate on the two qubits is  $\ket{\phi}=\hat{CNOT}\left(\ket{\psi}\otimes\ket{0}\right)=\alpha \ket{00}+\beta\ket{11}$ and evaluating the reduced density matrix related to the state of the system qubit, one gets  
\begin{equation}
\hat{\rho}_{ \text{fin}}=\left[\begin{matrix}
            |\alpha|^2 & 0 \\
            0 & |\beta|^2
        \end{matrix}\right],
\end{equation}
which is  exactly the density matrix that one would obtain after  a projective measurement process. The ideal setup is depicted in Fig. \ref{scheme}, where each rotation produces the intermediate Rabi evolution between every measurement, implemented by the $CNOT$ gates.

\begin{figure}
\includegraphics[width=0.9\columnwidth]{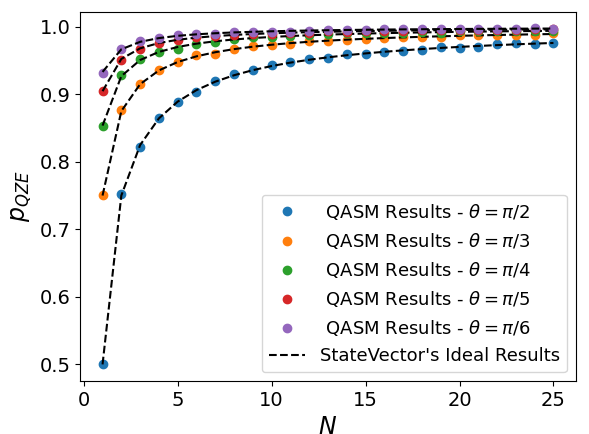}
\caption{Survival probability of an ideal qubit in the initial state, i.e. $\ket{0}$, depending on the number of intermediate measurements applied during the evolution at fixed $\theta$.}
\label{simulation}
\end{figure}

In the ideal case, imposing that  each $CNOT$ has a negligible runtime, the total evolution time can be fixed as $\theta=2\omega t$.  In Fig. \ref{simulation} we show the results of the simulations for different durations of the total evolution which are in agreement with the results shown in \cite{barik2020}. The ideal points are obtained through  Statevector simulations which produce for the final state the probability amplitudes of each element of the computational basis.
To reproduce the real experiment, simulations where performed also using QASM simulator backend from Qiskit Aer,  producing randomly distributed final measurement outputs. These simulations do not account for noise from decoherence, but only include statistical noise due to sampling. The weights of each component of the final state are thus reconstructed from the statistics of the outputs of the same repeated experiments. In order to have an accurate statistical description, we ran each quantum circuit $20000$ times.  As expected, for each $\theta$, the survival probability in the initial state  increases towards one as the number of intermediate measurements increases as well, freezing the Rabi oscillations. 

In this ideal case, the Rabi dynamics is assumed to be dominant with respect to any other dissipative effect. A real qubit is affected by decoherence processes due to the interaction with the external environment. These are the so called relaxation and dephasing: the former is associated to the spontaneous decay process from the excited state to the ground one in a non-unitary way and it has a characteristic time $T_1$, while the latter is related to the phenomenon which leads from a pure state to a mixed one, by suppressing the non-diagonal elements of the density matrix, and it is characterized by a specific $T_2$. It is important to highlight that, for the qubits of the devices provided by the IBM Quantum Experience, $T_1$ and $T_2$ are generically of tens or hundreds $\mu s$ in order of magnitude, and that the different gates are nothing but pulses with precise running times ($\langle \Delta t_{\text{single-qubit gates}}\rangle \simeq 35.7ns$ and $\langle\Delta t_{CNOT} \rangle\simeq 300ns $). Therefore, in the implementation on the real device, we expect that  increasing the number of operations, the decoherence becomes relevant.

The second case that we considered is the natural decay of the qubit. In the NISQ devices, for each qubit only one of the two computational states is stable, say $\ket{0}$, so after initializing the qubit on $\ket{1}$ we expect a relaxation to the ground state $\ket{0}$ due to the noise. The survival probability in the initial state is
\begin{equation}
p \left(t \right)= \braket{1|\hat{\rho} \left( t \right)|1}.
\end{equation}
Also in this case, after initializing the qubit  in its excited state, we try to freeze the spontaneous decay towards $\ket{0}$ introducing repeated measurements. As in the previous case, the measurements are implemented by ancillary qubits and $CNOT$ gates. In order to observe the QZE, we must consider the two decoherence effects which become significant around $T_1$ and $T_2$, imposing as a condition that the observation time $t$ should be such that $t\ll T_2\ll T_1$  \cite{matsuzaki2010}. The estimation of the observation time was done as follows. After the choice of a specific qubit on a certain device, we fixed a  number of intermediate measurements and the total time  $t$, which is given by the running times of all the gates employed,  the readout time (all these data are accessible on IBM Quantum Experience) and some intermediate temporal  delays.

\section{Implementations on a real device}
\label{results}

In this section we report  the  results of the experiments carried out  on an  IBM Quantum Experience real device. 
The implementation on a real quantum computer  is naturally affected by the noise which is not  negligible. The application of gates produces errors related to the discretization of their continuous parameters, but it can also introduce additional noise, especially when more qubits are involved, since a multi-qubits gate causes an interaction among qubits which enhances the decoherence effects.

\begin{figure}
\centering
\includegraphics[width=0.45\columnwidth]{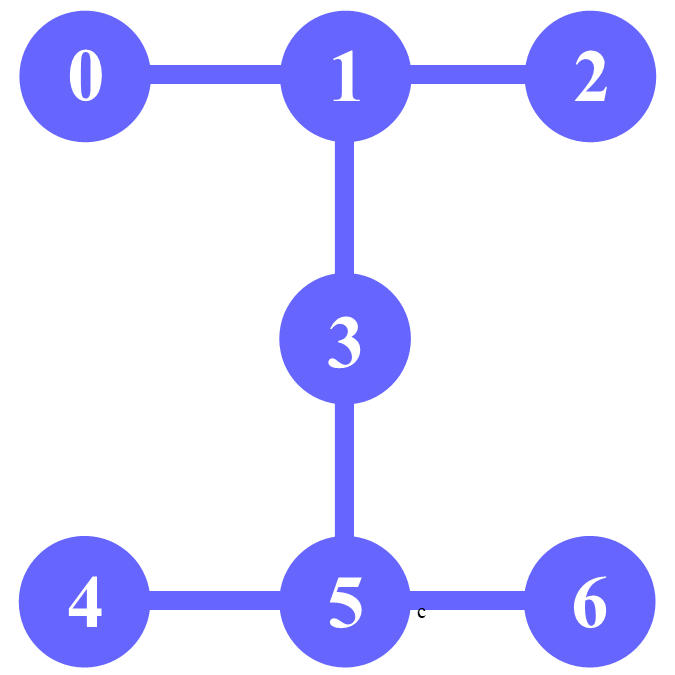}
\includegraphics[width=0.45\columnwidth]{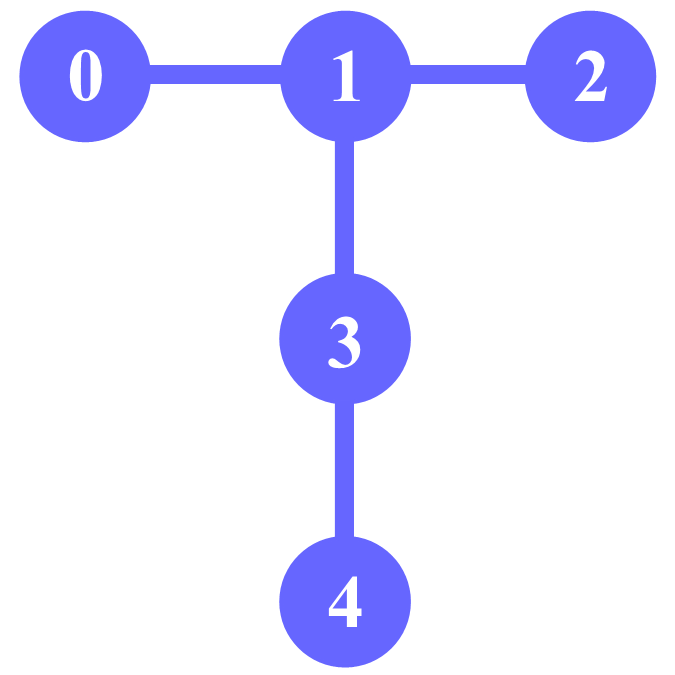}
\caption{ Coupling maps of the devices \texttt{ibm\_nairobi}  and \texttt{ibm\_perth} (on the left), and  \texttt{ibmq\_lima} (on the right). Each number is associated to a certain physical qubit, and every map describes how the different systems interact among them.}
\label{coupnairobi}
\end{figure}

Moreover,  the specific design of the quantum computer must be considered. Real devices are characterized by a specific coupling map which describes the physical connections between qubits. Since two-qubits gates can be directly applied only on connected qubits, 
the implementation on the real device must be  adapted accordingly. 
In Fig. \ref{coupnairobi} is reported the the coupling map of the devices  \texttt{ibm\_nairobi}, \texttt{ibm\_perth} and  \texttt{ibmq\_lima}. To simulate the Rabi dynamics of a qubit, we employed the devices \texttt{ibm\_nairobi} and \texttt{ibm\_osaka}, the coupling map of which is shown in App. \ref{osaka}, Fig. \ref{osaka16}.
In the former case, we considered the evolution of the physical qubit number $5$,  being the most connected and therefore suitable to be coupled directly to qubits  $3$, $4$ and $6$ by $CNOT$ gates, while in the latter case, for similar reasons, the qubit under study was the physical qubit number $39$.  This allowed us to reduce the time of execution, reducing the occurrence of the decoherence.  

\begin{figure*}
\includegraphics[width=\textwidth]{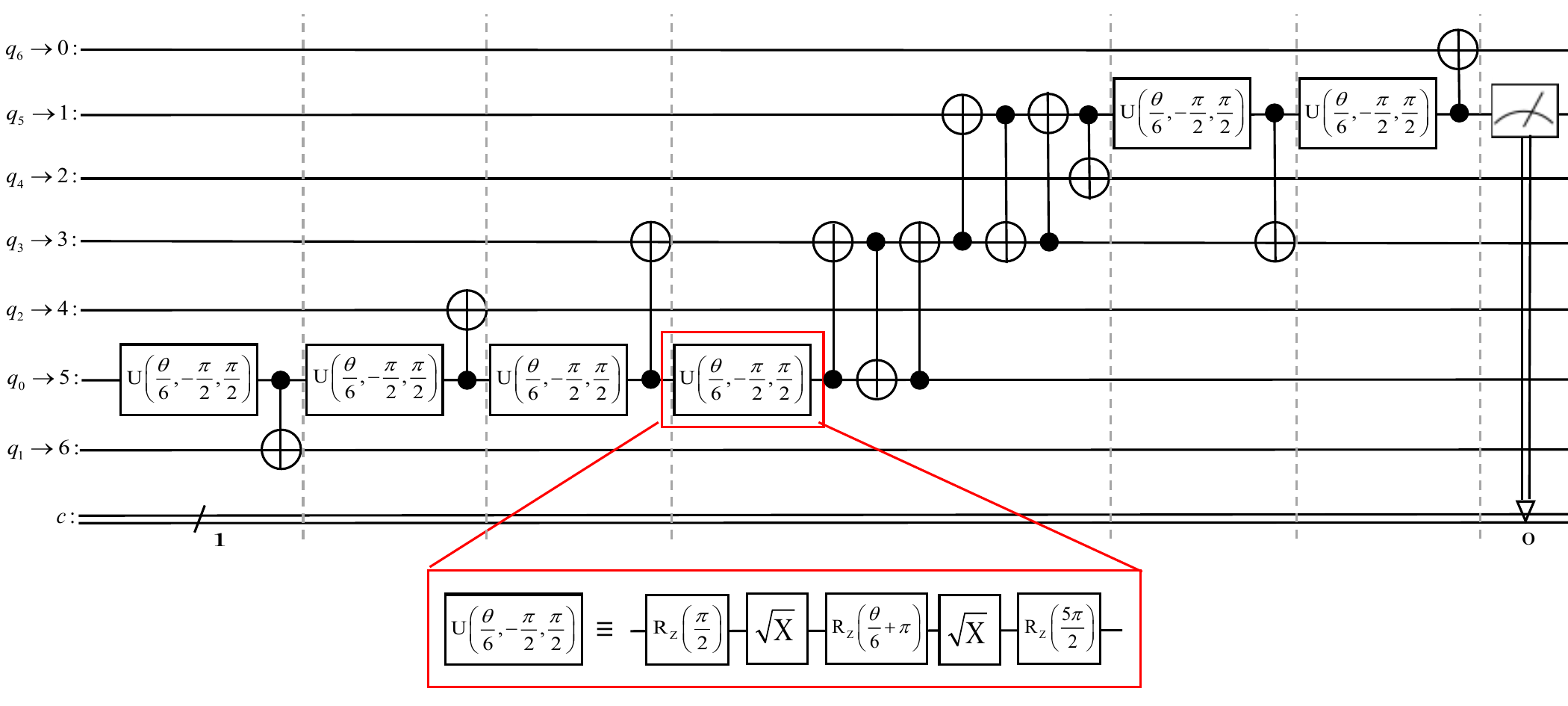}
\caption{ Circuit for the realization of the experiment with Rabi evolution on the real device \texttt{ibm\_nairobi}  starting from qubit $5$. The implementation of the circuit is reorganized taking into account the constraints of the device, i.e. the coupling map showed in Fig. \ref{coupnairobi} and the gates which can be actually realized. In fact, every machine can perform only a reduced set of simple specific gates that can be used to construct all the others, as we can see with the $U$ gate in the inset. To produce interaction between qubits which can not directly communicate, it is possible to apply three consecutive $CNOT$ gates alternating target and control, to get a $SWAP$ gate.}
\label{ex2}
\end{figure*}

Focusing on \texttt{ibm\_nairobi}, in order to simulate  the Rabi dynamics we prepared the circuit depicted in Fig.  \ref{ex2}  
with maximum level of optimization, 
reorganizing the order of the qubits accordingly to the  coupling map. 
Notice that, while Qiskit allows the implementation of the arbitrary single-qubit rotation $\hat{U}(\theta, \phi, \lambda)$  in Eq. (\ref{u}),  the real hardware has limitations that require adapting theoretical circuits. Each IBM device can only execute a reduced set of specific gates. For example, basis gates often include $R_z, \sqrt{X}$ and $CX$ ($CNOT$). The decomposition shown in Fig. \ref{ex2} reflects these hardware limitations. The coupling map of the device, as shown in Fig. \ref{coupnairobi}, indicates which qubits can directly interact with each other. The more complex construction shown in Fig. \ref{ex2}  ensures that all operations can be executed using the gates supported by the hardware. The decomposition of the operator $U(\frac{\theta}{6}, -\frac{\pi}{2}, \frac{\pi}{2})$ into $R_z$ and $\sqrt{X}$ reflects the need to use the basis gates of the device \cite{transpile}.

To produce interactions between qubits that cannot directly
communicate, it is necessary to use sequences of $SWAP$ gates, which
can be implemented by three consecutive $CNOT$ gates.  
The $SWAP$ gate can be represented, in the computational basis of the Hilbert space for two qubits, as 
\[
SWAP\doteq\left[
\begin{matrix}
    1 & 0 & 0 & 0\\
    0 & 0 & 1 & 0\\
    0 & 1 & 0 & 0\\
    0 & 0 & 0 & 1
\end{matrix}
\right],
\]
and it acts swapping the states of two qubits, i.e. $SWAP(\ket{\phi_1}\otimes\ket{\phi_2})=\ket{\phi_2}\otimes\ket{\phi_1}$. The $SWAP$ gate is frequently used in hardware where not all connections between qubits are possible to allow linking non-adjacent qubits \cite{swap}. The $SWAP$ is not a gate that IBM Quantum devices can directly implement. Given two qubits (sub)system and using three $CNOT$ gates which alternate the target qubit and the control one, we reproduce the same effect of the $SWAP$ gate. However, the construction of the $SWAP$ gate is not unique, since we can choose to start with either the first qubit or the second one as control. Every $CNOT$ gate has a specific running time which depends on both the qubits involved and on which one is the control (or the target). As a consequence, in  implementing a $SWAP$ operation one must consider the configuration of $CNOT$ gates which required less time to be executed.  Depending on the topology of the device, in order to operate a $CNOT$ between two non-adjacent qubits, one has to move the state of the qubit we are interested in until it becomes the state of a qubit which is adjacent to one designed as target.

\begin{figure}[H]
\includegraphics[width=0.9\columnwidth]{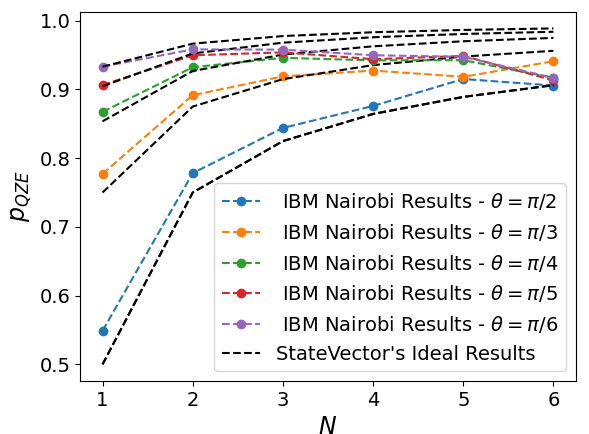}
\caption{The survival probability in the $\ket{0}$ state of the qubit number $5$ belonging to the device \texttt{ibm\_nairobi} as a function of the number of intermediate measurements introduced during the evolution at fixed $\theta$ in comparison with  the ideal results evaluated by \texttt{statevector\_simulator}.}
\label{real}
\end{figure}

The survival probability has been measured for different numbers of intermediate measurements and the results,  reported in Fig.  \ref{real}, are
qualitatively consistent with the ideal ones. There is also a quantitative good agreement, which is however limited from decoherence: in fact, in addition to statistical fluctuations, we can notice that the survival probability does not saturate to one but, with the increasing of the number of intermediate measurements, it tends to  underestimate the expected values.

\begin{table*} 
\begin{tabularx}{0.638\textwidth}{|c|c|c|c|c|}
    \hline \hline
    \textbf{Device} & \textbf{Qubit} & \textbf{Date and Time} & \textbf{Obs Time $[\mu s]$} & \textbf{$T[\mu s]$}  \\ 
    \hline \hline
    \texttt{ibm\_perth} & qr$[0]$ & May 15, 2023 2:08 AM & $2.667\pm 0.011$ & $5.61\pm 0.06$ \\
    \texttt{ibm\_nairobi} & qr$[2]$ & May 13, 2023 1:07 PM & $10.25\pm 0.02$ & $15.8\pm 1.1$ \\
    \texttt{ibm\_nairobi} & qr$[2]$ & May 15, 2023 2:26 AM & $10.25\pm 0.02$ & $15.8\pm 0.5$ \\
    \texttt{ibmq\_lima} & qr$[4]$ & May 13, 2023 2:06 AM & $10.48\pm 0.02$ & $16.0\pm0.4$\\
    \texttt{ibmq\_lima} & qr$[4]$ & Sep 10, 2023 11:39 PM & $10.48\pm 0.02$ & $17.5\pm0.5$\\
    \texttt{ibmq\_lima} & qr$[4]$ & Sep 10, 2023 5:10 PM & $8.380\pm 0.014$ & $14.5\pm0.4$\\
    \hline
\end{tabularx}
 \caption{Evaluation of the Zeno time $T$ for  different devices and qubits. In the last column we show the value of the Zeno time obtained by comparing the experimental data with the theoretical function $p_{QZE}(t)=\left[1-\left(\frac{t}{NT}\right)^2\right]^N$, where $N$ is the fixed maximum number of implicit measurements and $t$ is the evolution time.} 
\label{t1}
\end{table*}

\begin{figure*}
\includegraphics[width=\textwidth]{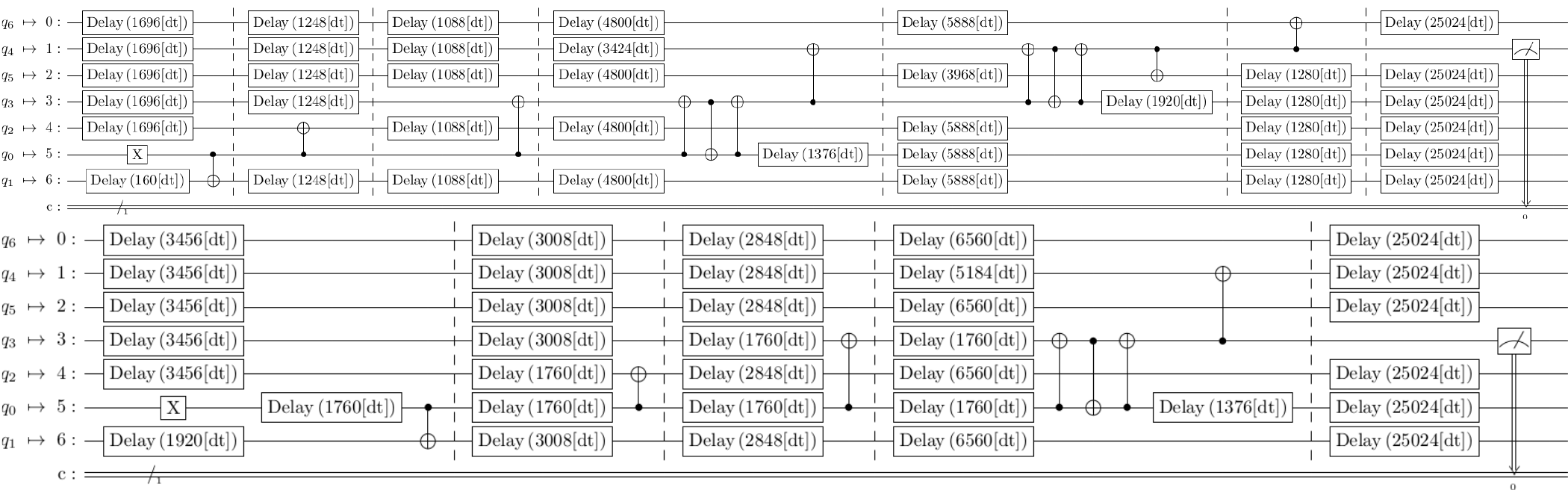}
\caption{ Circuits for the observation of the QZE during the spontaneous decay from $\ket{1}$ to  $\ket{0}$ of the state of qubit number $5$ of \texttt{ibm\_nairobi} with respectively six and four intermediate measurements. The running time of the former is $41024dt$, while the running time of the latter is  $40896dt$ ($1dt=0.\overline{2}ns$ in double precision). Three successive $CNOT$ gates disposed as shown in here realizes a $SWAP$ gate, whose effect is to exchange the states of the qubits involved. This is a necessary step when we want to entangle the states of two qubits that can not directly interact due to the coupling map of the device.}
\label{delaies}
\end{figure*}

Due to the unavoidable coupling of the system with the environment, the evolution of the qubit is affected by relaxation and dephasing. As a consequence, we expect to observe on the real device the QZE also without imposing any dynamics, such as the Rabi one, but simply initializing the state of the qubit in the excited one and leaving it free to evolve.  We stated in Section \ref{evolution} that the condition to have a quadratic behavior of the survival probability  is $t\ll T_2\ll T_1$, where $t$ is the observation time and $T_1$, $T_2$ are the decoherence times respectively for the relaxation and the dephasing. Taking into account  that $T_1$ and $T_2$ could change daily, or even hourly, we performed at different times and days the experiment for all the qubits of all the machines available on the IBM Quantum Experience platform until the $10^{th}$ of September $2023$. 
An example of the implemented quantum circuits with the employment of delays for the observation of the QZE with the state of a qubit which freely evolves is shown in Fig. \ref{delaies} and refers to the device \texttt{ibm\_nairobi}.

\begin{figure}
\centering
\includegraphics[width=0.9\columnwidth]{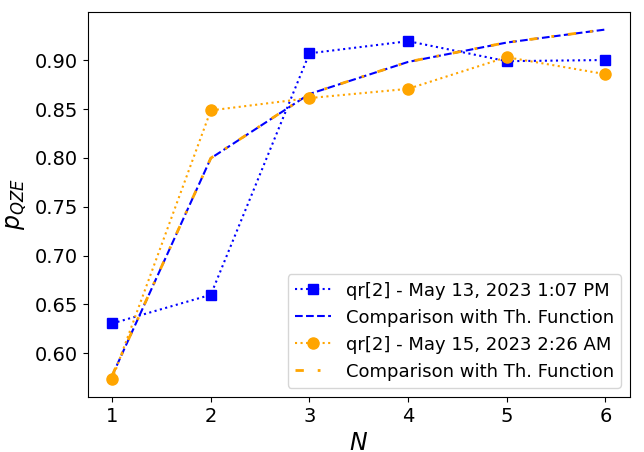}
\caption{ The evolution of the state of qubit number $2$ of the device \texttt{ibm\_nairobi} fixing the maximum number $N$ of intermediate measurements at $N=6$. The blue dashed line corresponds to $T=(15.8\pm1.1)\mu s$, while the orange one to $T=(15.8\pm0.5)\mu s$. }
\label{lima}
\end{figure}

\begin{figure}
\centering
\includegraphics[width=0.9\columnwidth]{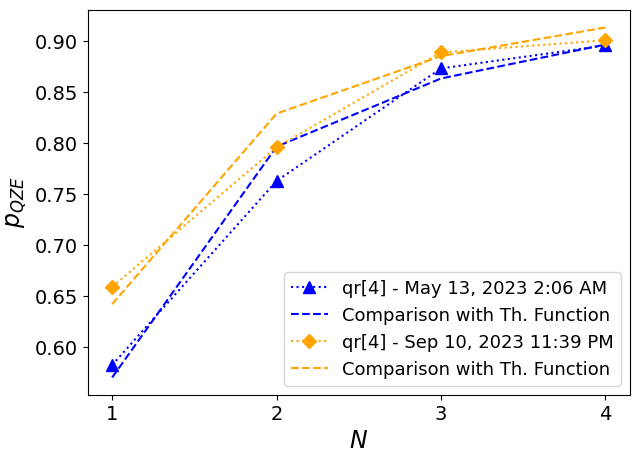}
\caption{The evolution of the state of qubit number $4$ of the device \texttt{ibmq\_lima} fixing the maximum number $N$ of intermediate measurements at $N=4$. The blue dashed line corresponds to $T=(16.0\pm0.4)\mu s$, while the orange one to $T=(17.5\pm0.5)\mu s$.}
\label{perth}
\end{figure}

\begin{figure}
\centering
\includegraphics[width=0.9\columnwidth]{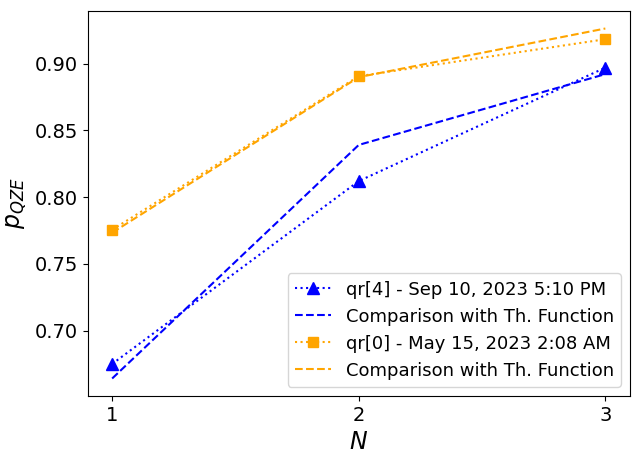}
\caption{ Evolution of the state of qubit number $4$ of the device \texttt{ibmq\_lima} (blue lines), compared with the theory with  $T=(14.5\pm0.4)\mu s$ (dashed line). Evolution of the state of qubit number $0$ of the device \texttt{ibm\_perth} (orange lines), compared with the theory with $T=(5.61\pm0.06)\mu s$ (dashed line). In both cases, the maximum number $N$ of intermediate measurements is fixed at $N=3$.}
\label{nairobi}
\end{figure}

In Tab. \ref{t1} 
are reported the data related to the clearest outputs obtained from the devices  \texttt{ibm\_nairobi},  \texttt{ibmq\_lima} and \texttt{ibm\_perth,} and they are plotted in in Figs. \ref{lima}, \ref{perth}, \ref{nairobi} in comparison with the decay function  
\begin{equation}
p_{QZE}(t)=\left(1-\frac{t^2}{N^2T^2}\right)^N,
\label{zeno}
\end{equation}
where $T$ is  the corresponding Zeno time.

The Eq. \eqref{zeno} can be easily obtained  \cite{BCS} following the same steps showed in Eq. \eqref{e4} considering a generic time-independent Hamiltonian $\hat{H}$ 
and $\ket{\psi,0}$ as the initial state of the system. The survival probability $p(t)$, that is, the probability to find the system in the same state $\ket{\psi,0}$ at time $t$, is given by
\[
p(t)=\left|\langle\psi,0\big|\psi,t\rangle\right|^2=\left|\langle\psi,0\big|\exp\left(-\frac{i}{\hbar}\hat{H}t\right)\big|\psi,0\rangle\right|^2.
\]
A short time expansion yields a quadratic behavior
\[
\begin{split}
    p(t)&\simeq\left|\langle\psi,0\big|\left(\hat{\mathbb{I}}-\frac{i}{\hbar}\hat{H}t-\frac{1}{2\hbar^2}\hat{H}^2t^2\right)\big|\psi,0\rangle\right|^2=\\
    &\simeq 1-\frac{\left(\langle\psi,0\big|\hat{H}^2\big|\psi,0\rangle-\langle\psi,0\big|\hat{H}\big|\psi,0\rangle^2\right)t^2}{\hbar^2}=\\
&=1-\frac{t^2}{T^2},
\end{split}
\]
where
\begin{equation}
T^2=\frac{\hbar^2}{\bra{\psi,0}\hat{H}^2\ket{\psi,0}-\bra{\psi,0}\hat{H}\ket{\psi,0}^2},
\end{equation}
is the square of the so-called Zeno time. If $N$ projective measurements are performed at time intervals $\Delta t=\frac{t}{N}$, then the survival probability at time $t$ is
\[
\begin{split}
    p_{QZE}(t)=\left[p\left(\frac{t}{N}\right)\right]^N&\simeq \left(1-\frac{t^2}{N^2T^2}\right)^N\\
&\xrightarrow[N\gg 1]{} \exp\left(-\frac{t^2}{NT^2}\right).
\end{split}
\]

The experimental data qualitatively follow the behavior of Eq. \eqref{zeno} with some $T$. In Fig. \ref{lima} we observe a worse agreement with Eq. \eqref{zeno}. A possible explanation is that the more  the multiple measurements are, the longer is the total time of observation, and this enhances the decoherence effects.
The fluctuations among the different outputs related to the same device and to the same qubit are due to frequent processes of recalibration of the decoherence times of the different systems.

Furthermore, we have to consider that, according also to Fig. \ref{delaies}, the measurements are not accurately equally-spaced in time, given the introduction in some cases of the $SWAP$ gate formalized through three adjacent $CNOT$ gates which are appropiately disposed. Nevertheless, what matters most in order to observe the QZE is to remain inside the temporal contraints imposed by the decoherence times, but this issue does not allow us to realize a proper fit with the function in Eq. \eqref{zeno} which was found instead for equally-spaced measurements.

Another detrimental element, always related to the $SWAP$ gate, is the fact that we can not make interact directly the totality of the qubits of a certain device. This implies the need for such a gate, which exchanges the states of two adjacent qubits, with a further enhancement of decoherence.

We notice that there is also an uncertainty on the evolution time $t$, as we can observe in Fig. \ref{delaies}. As we already said, we introduced some delays in order to compare experiments at fixed $t$. The evaluation of a delay depends on the temporal constraints imposed by the device we are working with. In fact, a physical device supports temporal intervals which are multiple of $dt$. In particular, $1dt=0.\overline{2}ns$ (in double precision) and it represents a discretization unit of temporal pulses sent as input to the machine. The granularity for the duration of a delay, that is the unit of time of which a delay should be multiple, is the least common multiple between the \texttt{pulse\_alignment} and the \texttt{acquire\_alignment}. In each analyzed case, we had $\text{\texttt{granularity}}=16 dt$. As a consequence, the delay communicated to the device is an integer multiple of such a value, corresponding to an approximation by excess of by defect of the time we really want to wait before the application of a certain gate. In the worst case, all the delays are approximated by excess (defect) and this leads to an error on the total evolution time which is equal to $N(16dt)$, where $N$ is the number of the total necessary delays, which coincides at most with the fixed maximum number of intermediate measurements.

All these issues lead to an estimation of different values of the Zeno time (but they all are of the order of $10\mu s$) and to a general tendency of the experimental data to underestimate the comparison function with the increase of the intermediate measurements.  Despite that,  the results of the experiments showed a QZE on trasmon qubits which freely evolve starting from the unstable state $\ket{1}$.

\section{Mitigation of SPAM errors} \label{ch:mitigation}
As a consequence of working on a NISQ computer,  all the results that we presented in the previous section are affected by noise. In this section we address  the SPAM errors which  are typically dominant in these devices. We focus only on the Rabi evolution since in the study of the free evolution the intention was to employ the QZE itself as a  mechanism to preserve the state against disturbances.

Since the real device \texttt{ibm\_nairobi} has been put out of commission after our first experiments, in order to  produce a mitigation of the SPAM errors we ran again the experiment of Rabi Oscillation on the device \texttt{FakeNairobiV2}.
The module \texttt{qiskit.providers.fake\_provider} contains fake providers and fake backends classes. The fake backends are built to mimic the behaviors of IBM Quantum systems using system snapshots. The system snapshots contain important information about the quantum system such as coupling map, basis gates, qubit properties ($T1$, $T2$, error rate, etc.) which are useful for testing the transpiler and performing noisy simulation of the system \cite{fakeprov,backendV2}.In general, the noise level of a fake backend reflects the parameters of the real device at the time of its
last operation or the latest update provided by IBM. It is important to note that IBM regularly updates
the parameters of fake backends to keep the simulation as accurate as possible compared to the physical
devices, even if the device is no longer operational.

Combining \texttt{complete\_meas\_cal} and \texttt{CompleteMeasFitter}, Qiskit returns the corresponding mitigation matrix \cite{calibrmatr} by calibration circuits where $2^M$ computational basis states are prepared and then measured ($M$ being the number of qubit used in the experiment). Readout mitigation is carried out first by measuring this mitigation matrix, and then its inverse is applied to the actual measurement distribution \cite{mitigation}. However, the preparation of computational states incurs errors, and these are not separately characterized by this Qiskit procedure. In order to take into account these errors, we resorted to the State Preparation and Readout Mitigation (SPRM) procedure described in \cite{mitigation}, and then compared the two mitigations.

\begin{figure}[H]
    \centering
    \begin{minipage}{0.9\columnwidth}
        \centering
        \includegraphics[width=\textwidth]{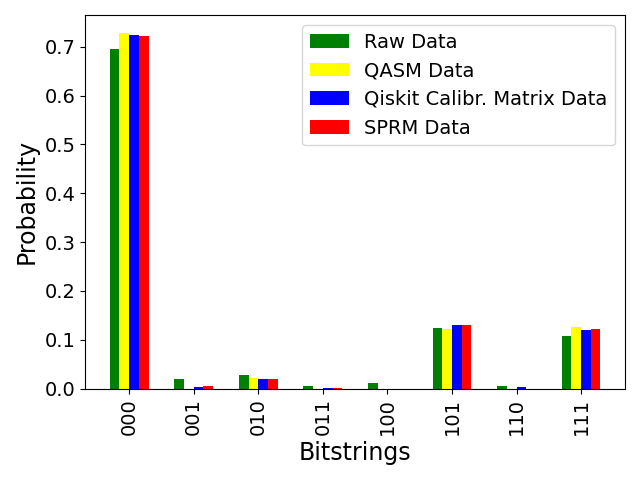}
    \end{minipage}
    \hfill
    \begin{minipage}{0.9\columnwidth}
        \centering
        \includegraphics[width=\textwidth]{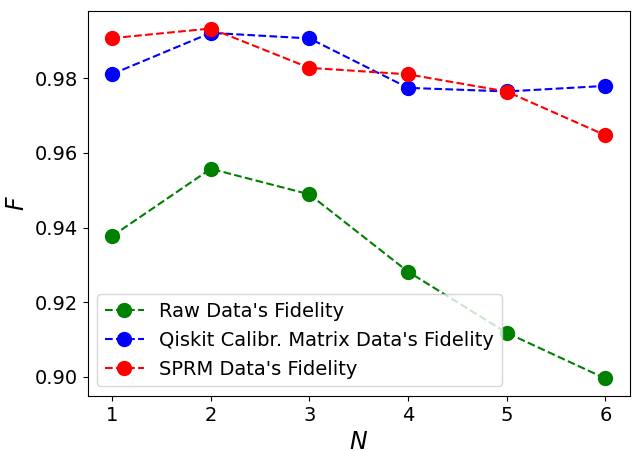}
    \end{minipage}
    \hfill
    \begin{minipage}{0.9\columnwidth}
        \centering
        \includegraphics[width=\textwidth]{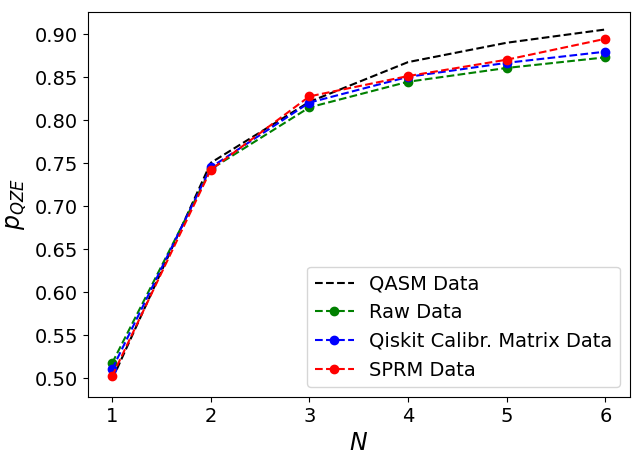}
    \end{minipage}
    \caption{Comparison between QASM, the raw data of the experiment, the Qiskit mitigated data and the SPRM mitigated data. Results for $\theta=\frac{\pi}{2}$. First panel: comparison between the  probability distribution of the outputs. Second panel:  fidelity between  experimental data (with and without mitigation) and QASM. Third panel: survival probability of the qubit in the initial state.  }
    \label{fig:one}
\end{figure}

We performed the experiment for   $\theta=\frac{\pi}{2}, \frac{\pi}{3}, \frac{\pi}{4}, \frac{\pi}{5}, \frac{\pi}{6}$. In Fig. \ref{fig:one} we report the results for   $\theta=\frac{\pi}{2}$, while the plots for the other angles, which are qualitatively similar, are included in App. \ref{extrarun}. In the first panel of  Fig. \ref{fig:one}  we show the comparison between  the probability distributions  (with two ancillary qubits) of the bitstrings output obtained from the raw experiment,  the QASM simulation, after the Qiskit calibration matrix mitigation   and after the SPRM procedure. 

In the second panel is reported the classical fidelity between the experimental probability distribution and  the exact one $\left\{\left( p_{QZE}^{(QASM)}\right)_i\right\}_{i=1}^{2^M-1}$ (obtained by QASM) defined as 
\begin{equation}
F = \left[\sum_{i=0}^{2^M-1}\sqrt{\left(p_{QZE}\right)_i\left(p_{QZE}^{(QASM)}\right)_i}\right]^2
\end{equation}
where $\left(p_{QZE}\right)_i$ is the probability 
of the state $\ket{i}$ (in base ten) \cite{mitigation}, and $M=N+1$.
In the last panel, we compare the survival probability of the qubit in the initial state  with and without mitigation.

Both the mitigation techniques improve the fidelity of the output probability distribution with respect to the raw one. Moreover, between the two mitigation techniques themselves we can observe that the corresponding fidelities at fixed number of intermediate measurements are comparable, but marginalizing with respect to the degrees of freedom of all the $N$ ancillary qubits and plotting the survival probability of the other one, we  observe in the third panel of Fig. \ref{fig:one} that we achieve a better mitigation with the SPRM method.

An analysis of the mitigated data executed on a currently working real device, i.e. \texttt{ibm\_osaka}, is reported in App. \ref{osaka}. Due to the limited resources at our disposal, we performed the experiment only for $\theta=\frac{\pi}{2}$.

\section{Conclusions} \label{ch:conclusion}
In this paper we reported  the observation of the QZE on  IBM NISQ devices. We studied  the survival probability of a single qubit in its initial state after the effect of multiple measurements. 

We first employed the quantum circuit to mimic a Rabi time-evolution, as first proposed by \cite{barik2020}, introducing a set of $CNOT$ gates to perform the intermediate measurements. After a first  simulation by QASM  we reproduced the experiment on a real device finding a net reduction of the oscillating rate, with a good qualitative agreement with the simulation.

We also tried to observe the QZE during the free evolution of the state of a qubit from the unstable $\ket{1}$ to $\ket{0}$. In order to observe the QZE, while in the previous case the time distance between two  subsequent measurements had to be  much smaller than the Rabi period, here it has to be  much smaller than the decoherence times. Unfortunately, we do not have a full control of the decoherence times, since they are continuously recalibrated,  and additionally each circuit is inserted in a queue that could be roughly long during which $T_1$ and $T_2$ may change. This is why in this case the observation of  the QZE turned out to be more demanding. 
Some qubits exhibit decoherence times with very little relative variations:  there are some qubits which more easily show the QZE, but for a more quantitative analysis of the reason why some qubits show this behavior rather than others it could be important to have precise information about the Hamiltonian of the single qubit under examination. The platform IBM Quantum Experience allows us to know formally the Hamiltonian of each device, but we could not be able to have access to more specific information about the parameters of those Hamiltonians. 

As it is shown in Section \ref{results}, we were able to observe the Quantum Zeno Effect  in both simulations and experiments.

Furthermore, SPAM errors are considered and error mitigations have been performed using both Qiskit calibration matrix  and SPRM procedure. Both routines provide significant improvements compared to the raw data confirming the expected QZE behavior.

{\bf Acknowledgements }-- The authors acknowledge the Ministero dell'Università e della Ricerca (MUR) and the Project PRIN 2022 number 2022W9W423 funded by the European Union  Next Generation EU; IBM, the IBM logo and ibm.com are trademarks of International Business Machines Corp., registered in many jurisdictions worldwide. Other product and service names might be trademarks of IBM or other companies. The current list of IBM trademarks is available (see \href{https://www.ibm.com/legal/copytrade}{https://www.ibm.com/legal/copytrade}).
 S. P. and C. C. acknowledge financial support from the University of L'Aquila by the internal project "Variational Quantum Eigensolver methods for the disordered Su-Schrieffer-Heeger model."
We acknowledge the CINECA award under the ISCRA initiative, for the availability of high-performance computing resources and support.\\

{\bf Data availability statement }-- The data that support the findings of this study are available upon request from the authors.

 \begin{appendix}

 \section{Mitigated data from \texttt{FakeNairobiV2}}
\label{extrarun}

In this appendix we report the plots related to SPAM error-mitigation for $\theta= \frac{\pi}{3}, \frac{\pi}{4}, \frac{\pi}{5}, \frac{\pi}{6}$. 

\begin{figure}[H]
    \centering
    \begin{minipage}{0.9\columnwidth}
        \centering
        \includegraphics[width=\textwidth]{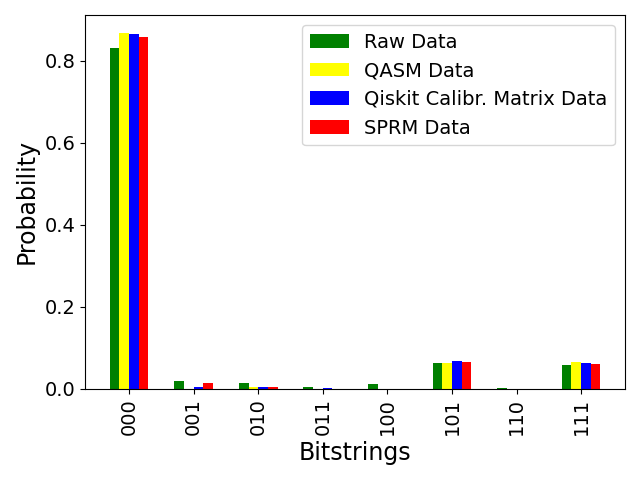}
    \end{minipage}
    \hfill
    \begin{minipage}{0.9\columnwidth}
        \centering
        \includegraphics[width=\textwidth]{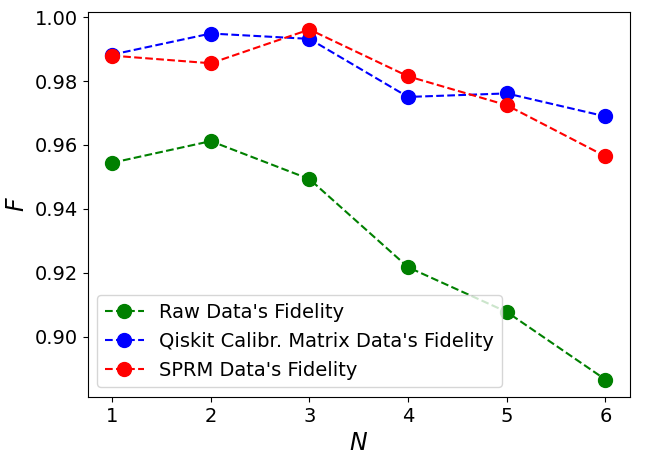}
    \end{minipage}
    \hfill
    \begin{minipage}{0.9\columnwidth}
        \centering
        \includegraphics[width=\textwidth]{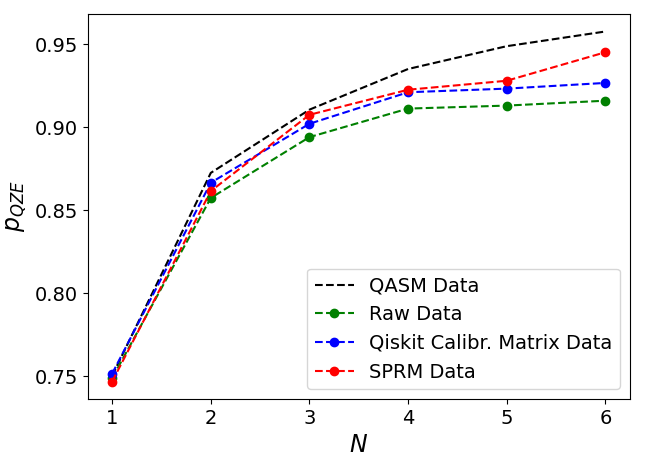}
    \end{minipage}
    \caption{Comparison between QASM, the raw data of the experiment, the Qiskit mitigated data and the SPRM mitigated data. Results for $\theta=\frac{\pi}{3}$. First panel: comparison between the  probability distribution of the outputs. Second panel:  fidelity between  experimental data (with and without mitigation) and QASM. Third panel: survival probability of the qubit in the initial state.}
    \label{fig:two}
\end{figure}

\begin{figure}[H]
    \centering
    \begin{minipage}{0.9\columnwidth}
        \centering
        \includegraphics[width=\textwidth]{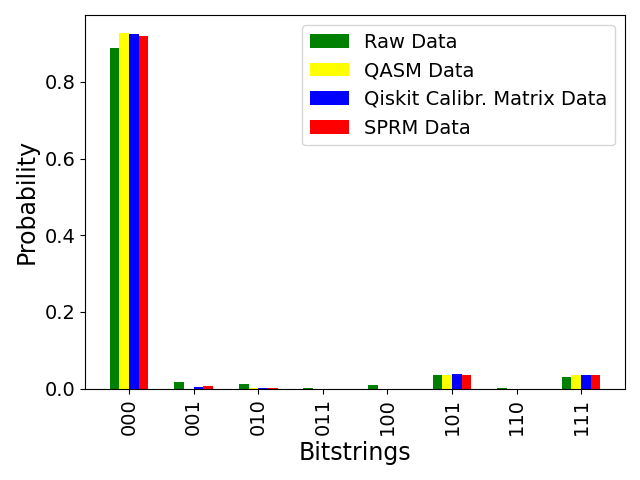}
    \end{minipage}
    \hfill
    \begin{minipage}{0.9\columnwidth}
        \centering
        \includegraphics[width=\textwidth]{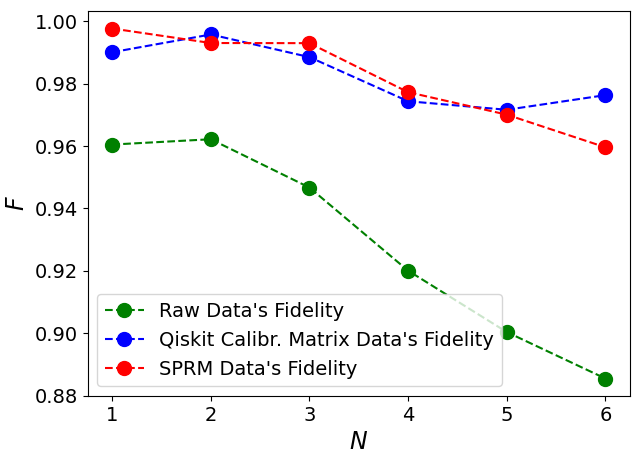}
    \end{minipage}
    \hfill
    \begin{minipage}{0.9\columnwidth}
        \centering
        \includegraphics[width=\textwidth]{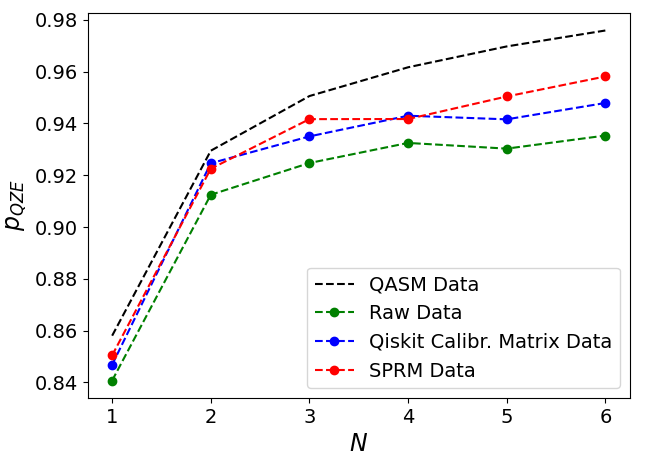}
    \end{minipage}
        \caption{Comparison between QASM, the raw data of the experiment, the Qiskit mitigated data and the SPRM mitigated data. Results for $\theta=\frac{\pi}{4}$. First panel: comparison between the  probability distribution of the outputs. Second panel:  fidelity between  experimental data (with and without mitigation) and QASM. Third panel: survival probability of the qubit in the initial state.}
    \label{fig:three}
\end{figure}

\begin{figure}[H]
    \centering
    \begin{minipage}{0.9\columnwidth}
        \centering
        \includegraphics[width=\textwidth]{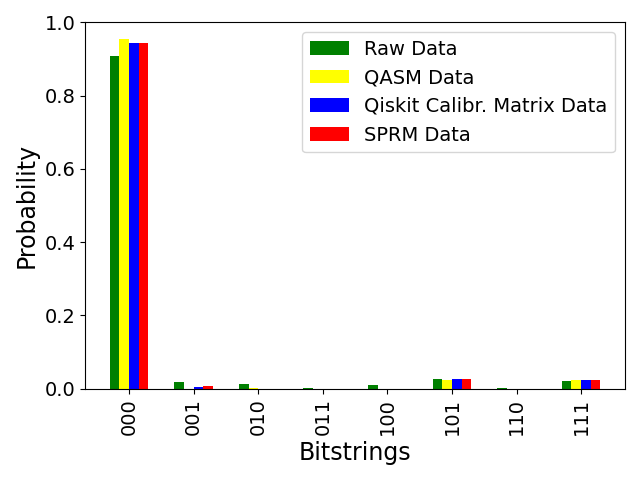}
    \end{minipage}
    \hfill
    \begin{minipage}{0.9\columnwidth}
        \centering
        \includegraphics[width=\textwidth]{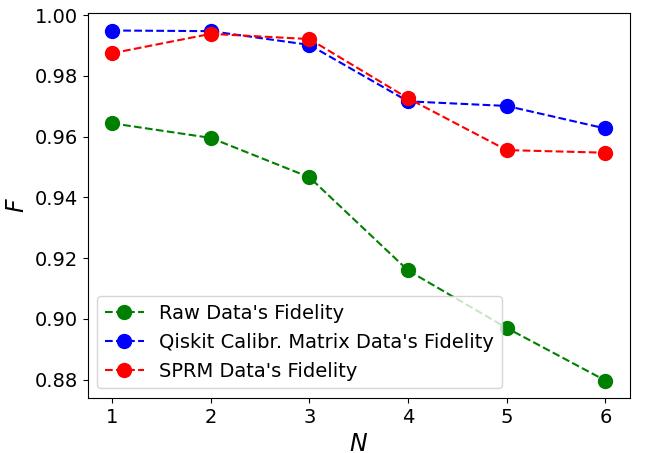}
    \end{minipage}
    \hfill
    \begin{minipage}{0.9\columnwidth}
        \centering
        \includegraphics[width=\textwidth]{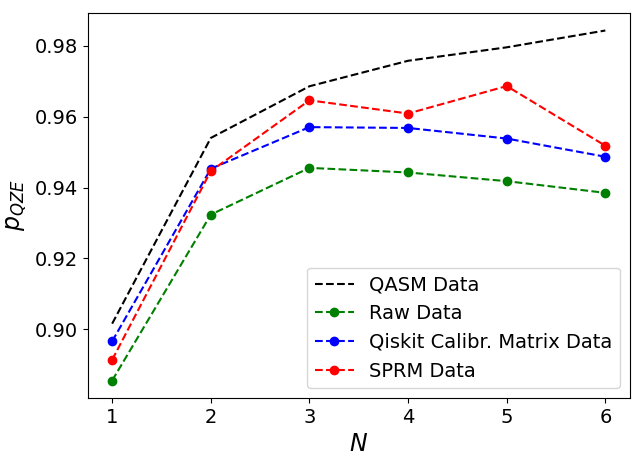}
    \end{minipage}
    \caption{Comparison between QASM, the raw data of the experiment, the Qiskit mitigated data and the SPRM mitigated data. Results for $\theta=\frac{\pi}{5}$. First panel: comparison between the  probability distribution of the outputs. Second panel:  fidelity between  experimental data (with and without mitigation) and QASM. Third panel: survival probability of the qubit in the initial state.}
    \label{fig:four}
\end{figure}
\begin{figure}[H]
    \centering
    \begin{minipage}{0.9\columnwidth}
        \centering
        \includegraphics[width=\textwidth]{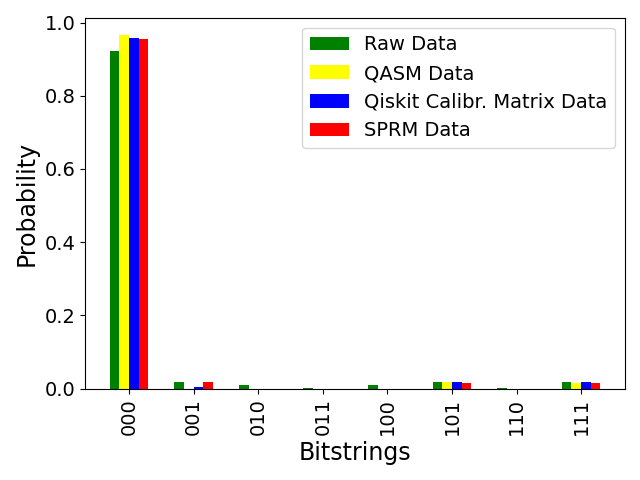}
    \end{minipage}
    \hfill
    \begin{minipage}{0.9\columnwidth}
        \centering
        \includegraphics[width=\textwidth]{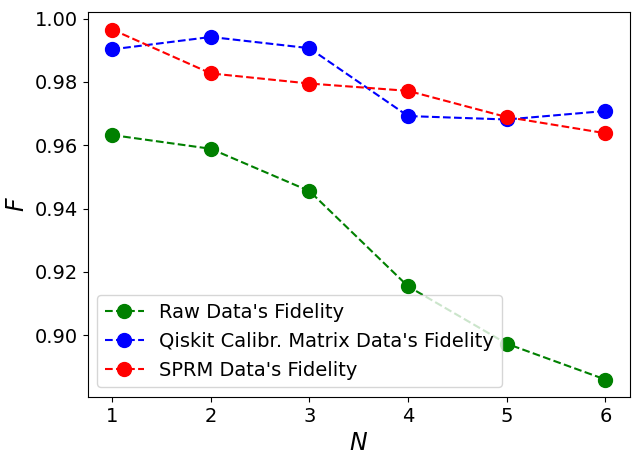}
    \end{minipage}
    \hfill
    \begin{minipage}{0.9\columnwidth}
        \centering
        \includegraphics[width=\textwidth]{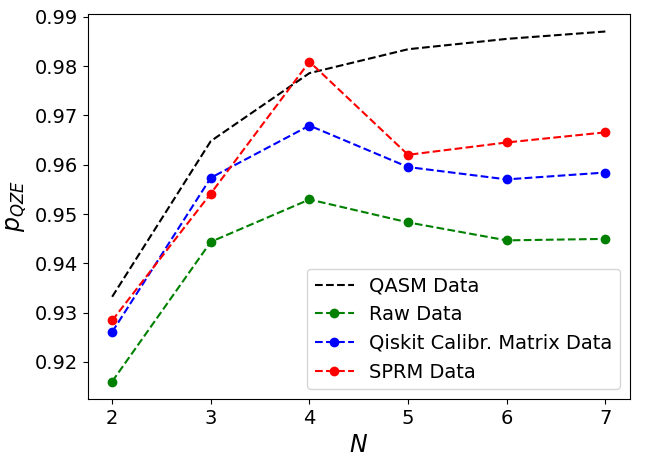}
    \end{minipage}
    \caption{Comparison between QASM, the raw data of the experiment, the Qiskit mitigated data and the SPRM mitigated data. Results for $\theta=\frac{\pi}{6}$. First panel: comparison between the  probability distribution of the outputs. Second panel:  fidelity between  experimental data (with and without mitigation) and QASM. Third panel: survival probability of the qubit in the initial state.}
    \label{fig:five}
\end{figure}

  \section{Mitigated data from \texttt{ibm\_osaka}}
 \label{osaka}

In this appendix we report some measurements from the device  \texttt{ibm\_osaka} which are more recent than the ones reported in the main text. Meanwhile the  IBM devices have changed, and the $CNOT$ gate does not belong anymore to the set of basis gates of the available devices. Nowadays, we find the $ECR$ (Echoed Cross-Resonance) gate, which is a maximally entangling gate and is equivalent to a $CNOT$ up to a single-qubit pre-rotations. The echoing procedure mitigates some unwanted terms to cancel in an experiment \cite{ECR}. 

 \begin{figure}[H]
    \centering
    \begin{minipage}{0.9\columnwidth}
        \centering
        \includegraphics[width=\textwidth]{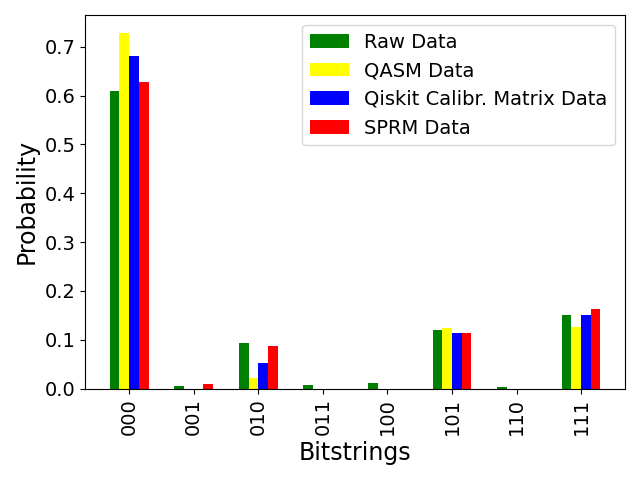}
    \end{minipage}

    \vskip\baselineskip

    \begin{minipage}{0.9\columnwidth}
        \centering
        \includegraphics[width=\textwidth]{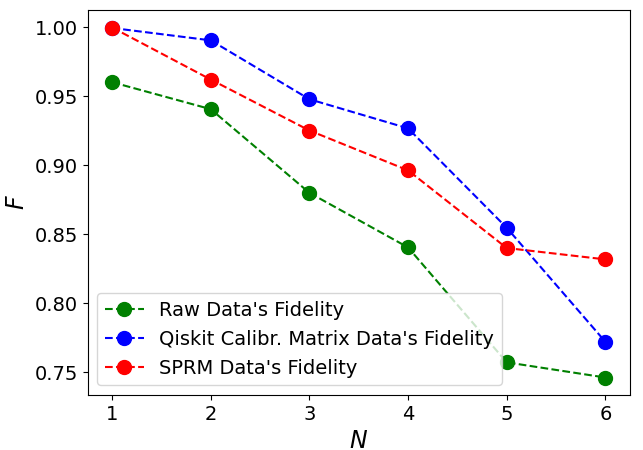}
    \end{minipage}
    \hfill
    \begin{minipage}{0.9\columnwidth}
        \centering
        \includegraphics[width=\textwidth]{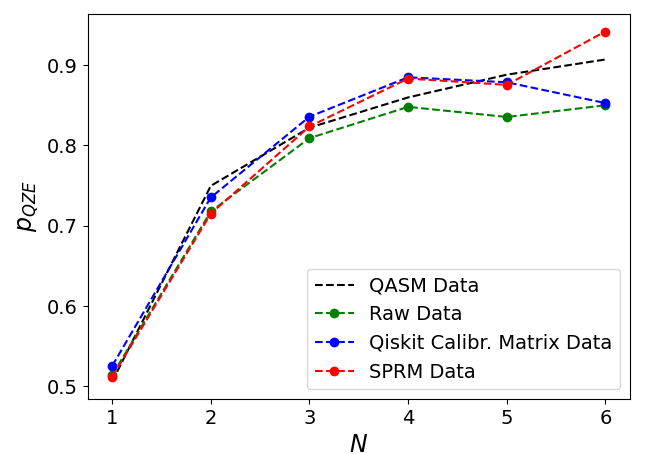}
    \end{minipage}
    \caption{ Comparison between QASM, the raw data of the experiment, the Qiskit mitigated data and the SPRM mitigated data from the device  \texttt{ibm\_osaka}. Results for $\theta=\frac{\pi}{2}$. First panel: comparison between the  probability distribution of the outputs. Second panel:  fidelity between  experimental data (with and without mitigation) and QASM. Third panel: survival probability of the qubit in the initial state.}
    \label{fig:six}
\end{figure}

\begin{figure}[H]
    \centering
    \includegraphics[width=0.9\columnwidth]{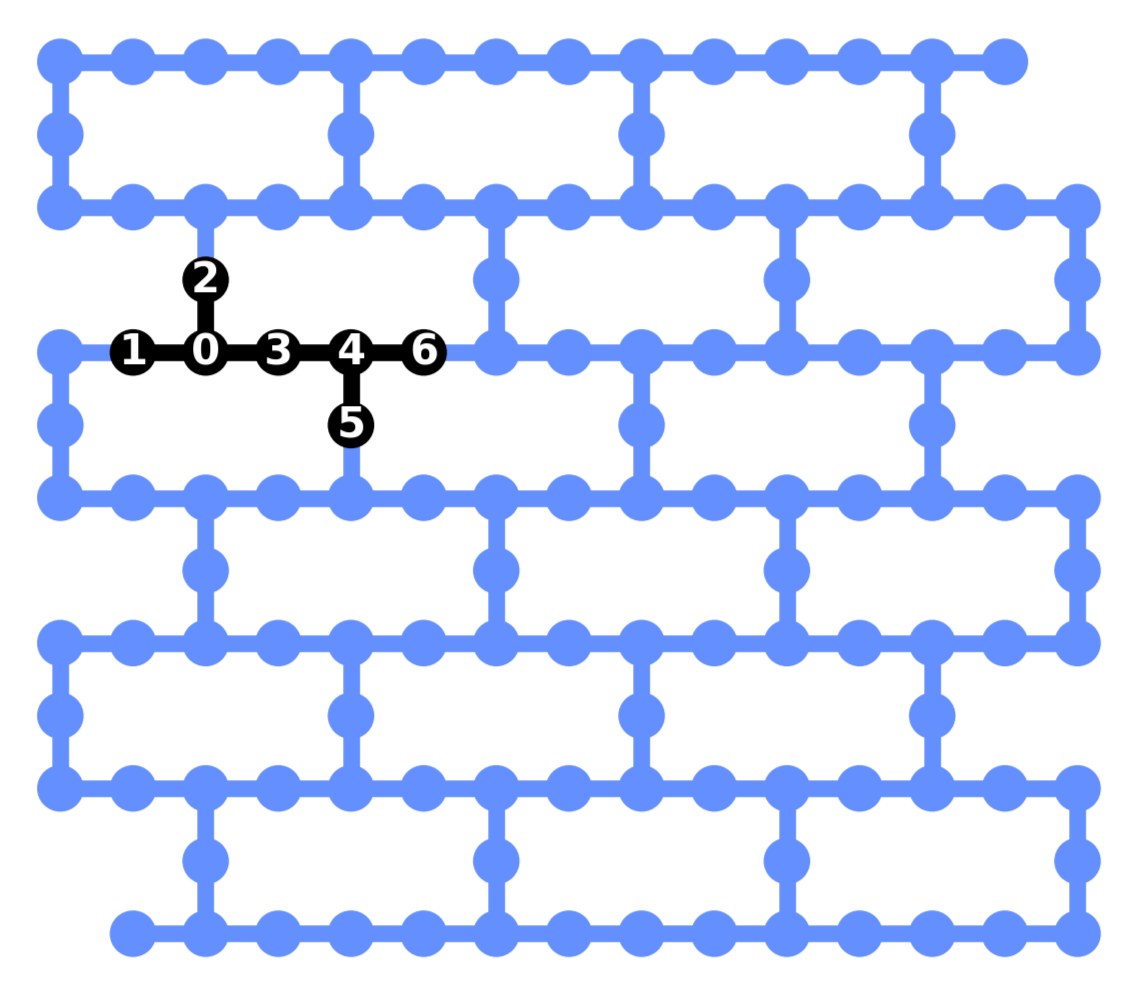}
    \caption{ Circuit layout of the device  \texttt{ibm\_osaka} with the virtual qubits we used and the corresponding labelling. }
\label{osaka16}
\end{figure}

 \end{appendix}


\begin{thebibliography}{XX}

\bibitem{degasperis1974}
A. Degasperis, L. Fonda andG. C. Ghirardi,
{\it Does the lifetime of an unstable system depend on the measuring apparatus?},
\href{https://doi.org/10.1007/BF02731351}{Nuovo Cimento A  {\bf 21}, 471 (1974)}.

\bibitem{misra1977}
B. Misra and E. C. G. Sudarshan,
{\it The Zeno's paradox in quantum theory},
\href{https://pubs.aip.org/aip/jmp/article-pdf/18/4/756/8148782/756\_1\_online.pdf}{J. Math. Phys. {\bf 18}, 756 (1977)}.

\bibitem{chiu1977}
C. B. Chiu, E. C. G. Sudarshan and B. Misra,
{\it The time evolution of unstable quantum states and a resolution of Zeno's paradox},
\href{https://link.aps.org/doi/10.1103/PhysRevD.16.520}{Phys. Rev. D. {\bf16},  520 (1977)}.

\bibitem{joos1984}
E. Joos,
{\it Continuous measurement: Watchdog effect versus golden rule},
\href{https://https://doi.org/10.1103/PhysRevD.29.1626}{Phys. Rev. D {\bf 29}, 1626 (1974)}.

\bibitem{cook1988}
Richard J. Cook,
{\it What are Quantum Jumps?},
\href{https://dx.doi.org/10.1088/0031-8949/1988/T21/009}{Phys. Scr.  {\bf1988(T21)} (1988)}.

\bibitem{nakazato1996}
H. Nakazato, M. Namiki, S. Pascazio and H. Rauch ,
{\it Understanding the quantum Zeno effect}, 
\href{https://doi.org/10.1016/0375-9601(96)00350-7 }{Phys. Lett. A {\bf 217}, 203 (1996)}.

\bibitem{presilla1996}
C. Presilla, R.Onofrio and U. Tambini
{\it Measurement Quantum Mechanics and Experiments on Quantum Zeno Effect}
\href{https://doi.org/10.1006/aphy.1996.0052}
{Ann. Physics {\bf 248}, 95121 (1996)}

\bibitem{facchi2009}
P. Facchi, G. Marmo and S. Pascazio,
{\it Quantum Zeno dynamics and quantum Zeno subspaces},
\href{https://iopscience.iop.org/article/10.1088/1742-6596/196/1/012017}{J. Phys.: Conf. Ser. {\bf196}, 012017 (2009)}.

\bibitem{pascazio2014}
S. Pascazio,
{\it All You Ever Wanted to Know About the Quantum Zeno Effect in 70 Minutes}, 
\href{https://doi.org/10.1142/S1230161214400071}{Open Sys. Inf. Dyn. {\bf 21}, 1440007 (2014)}.

\bibitem{giacosa2017}
F. Giacosa,
{\it Non-exponential Decay in Quantum Field Theory and in Quantum Mechanics: The Case of Two (or More) Decay Channels},
\href{https://doi.org/10.1007/s10701-012-9667-3}{Found. Phys. {\bf 42}, 1262  (2012)}.

\bibitem{lucia2023}
U. Lucia,
{\it Irreversible and quantum thermodynamic considerations on the quantum Zeno effect},
\href{https://doi.org/10.1038/s41598-023-38040-w}{Sci. Rep.  {\bf 13}, 10763 (2023)}.

\bibitem{itano1990}
Wayne M. Itano, D. J. Heinzen, J. J. Bollinger and D. J. Wineland, 
{\it Quantum Zeno effect},
\href{https://link.aps.org/doi/10.1103/PhysRevA.41.2295}{Phys. Rev. A  {\bf41}, 2295 (1990)}.

\bibitem{kwiat1995}
P. Kwiat, H. Weinfurter, T. Herzog, A. Zeilinger, M. A. Kasevich,
{\it Interaction-Free Measurement},
\href{https://link.aps.org/doi/10.1103/PhysRevLett.74.4763}{Phys. Rev. Lett. {\bf 74},  4763 (1995)}.

\bibitem{nagels1997}
B. Nagels, L. J. F. Hermans, P. L. Chapovsky, 
{\it Quantum Zeno Effect Induced by Collisions},
\href{https://link.aps.org/doi/10.1103/PhysRevLett.79.3097}{Phys. Rev. Lett. {\bf 79},  3097 (1997)}.

\bibitem{drewsen2000}
K. MÃžlhave and M. Drewsen, 
{\it Demonstration of the continuous quantum Zeno effect in optical pumping},
\href{https://www.sciencedirect.com/science/article/pii/S0375960100001663}{Phys. Lett. A {\bf 268}, 45 (2000)}.

\bibitem{fischer2001}
M. C. Fischer, B. Guti\'errez-Medina, M. G. Raizen, 
{\it Observation of the Quantum Zeno and Anti-Zeno Effects in an Unstable System},
\href{https://link.aps.org/doi/10.1103/PhysRevLett.87.040402}{Phys. Rev. Lett. {\bf 87}, 040402 (2001)}.

\bibitem{zhu2014}
B. Zhu, B. Gadway, M. Foss-Feig, J. Schachenmayer, M. L. Wall, K. R. A. Hazzard, B. Yan, S. A. Moses, J. P. Covey, D. S. Jin, J. Ye, M. Holland, A. M. Rey,
{\it Suppressing the Loss of Ultracold Molecules Via the Continuous Quantum Zeno Effect},
\href{https://link.aps.org/doi/10.1103/PhysRevLett.112.070404}{Phys. Rev. Lett. {\bf 112},  070404 (2014)}.

\bibitem{facchi2002}
P. Facchi, S. Pascazio,
{\it Quantum Zeno Subspaces},
\href{https://link.aps.org/doi/10.1103/PhysRevLett.89.080401}{Phys. Rev. Lett. {\bf 89}, 080401 (2002)}.

\bibitem{facchi2004}
P. Facchi, D. A. Lidar, S. Pascazio,
{\it Unification of dynamical decoupling and the quantum Zeno effect},
\href{https://link.aps.org/doi/10.1103/PhysRevA.69.032314}{Phys. Rev. A {\bf69}, 032314 (2004)}.

\bibitem{facchi2006}
P. Facchi, R. Fazio, G. Florio, S. Pascazio, T. Yoneda,
{\it Zeno Subspaces for Coupled Superconducting Qubits},
\href{https://doi.org/10.1007/s10701-005-9033-9}{Found. Phys. {\bf 36}, 500 (2006)}.

\bibitem{facchi2010}
P. Facchi, M. Ligab\`o
{\it Quantum Zeno effect and dynamics},
\href{https://doi.org/10.1063/1.3290971}{J. Math. Phys. {\bf 51}, 022103 (2010)}.

\bibitem{martin2018}
S. Hacohen-Gourgy, L. P. Garc\'{\i}a-Pintos, L. S. Martin, J. Dressel, I. Siddiqi,
{\it Incoherent Qubit Control Using the Quantum Zeno Effect},
\href{https://link.aps.org/doi/10.1103/PhysRevLett.120.020505}{Phys. Rev. Lett. {\bf 120},  020505 (2018)}.

\bibitem{facchi2005}
P. Facchi, S. Tasaki, S. Pascazio, H. Nakazato, A. Tokuse, D. A. Lidar,
{\it Control of decoherence: Analysis and comparison of three different strategies},
\href{https://link.aps.org/doi/10.1103/PhysRevA.71.022302}{Phys. Rev. A {\bf 71}, 022302 (2005)}.

\bibitem{paz2012zeno}
G. Paz-Silva, A. T.  Rezakhani, J. M. Dominy and D. A. Lidar,
{\it Zeno effect for quantum computation and control},
\href{https://journals.aps.org/prl/abstract/10.1103/PhysRevLett.108.080501}{Phys. Rev. Lett. {\bf 108}, 080501 (2012)}

\bibitem{fonda1978}
L. Fonda, G. C. Ghirardi and  A. Rimini,
{\it Decay theory of unstable quantum systems},
\href{https://dx.doi.org/10.1088/0034-4885/41/4/003}{Rep. Prog. Phys., {\bf 41}, 587 (1978)}.

\bibitem{pascazio1996}
H. Nakazato, M. Namiki, S. Pascazio,
{\it Temporal behavior of quantum mechanical systems},
\href{https://doi.org/10.1142/S0217979296000118}{Int. J. Mod. Phys. B, {\bf 10}, 247 (1996)}.

\bibitem{kofman2001}
A. G. Kofman and G. Kurizki,
{\it Frequent Observations Accelerate Decay: The anti-Zeno Effect}, 
\href{https://doi.org/10.1515/zna-2001-0113}{Z. Naturforsch. A {\bf 56}, 83 (2001)}.

\bibitem{facchi2001}
P. Facchi, H. Nakazato, S. Pascazio,  
{\it From the Quantum Zeno to the Inverse Quantum Zeno Effect}, 
\href{https://link.aps.org/doi/10.1103/PhysRevLett.86.2699}{Phys. Rev. Lett. {\bf 86(13)}, 2699 (2001)}.

\bibitem{antoniou2001}
I. Antoniou, E. Karpov, G. Pronko, E. Yarevsky, 
{\it Quantum Zeno and anti-Zeno effects in the Friedrichs model},
\href{https://link.aps.org/doi/10.1103/PhysRevA.63.062110}{Phys. Rev. A {\bf 63(6)},  062110 (2001)}.

\bibitem{na2021}
K. Na,
{\it Revisiting quantum Zeno effect and anti-Zeno effect: Universality vs non-universality}, 
\href{https://doi.org/10.1063/5.0050473}{J. Math. Phys. {\bf 62}, 122101 (2021)}.

\bibitem{lewenstein2000}
M. Lewenstein, K. Rza\ifmmode \mbox{\c{}}\else \c{}\fi{}\ifmmode \dot{z}\else \.{z}\fi{}ewski,
{\it Quantum anti-Zeno effect},
\href{https://link.aps.org/doi/10.1103/PhysRevA.61.022105}{Phys. Rev. A {\bf 61}, 022105 (2000)}.

\bibitem{franson2004quantum}
J. D. Franson,B. C. Jacobs  and T. B. Pittman,
{\it Quantum computing using single photons and the Zeno effect},
\href{https://journals.aps.org/pra/abstract/10.1103/PhysRevA.70.062302}{Phys. Rev. A {\bf 70},  062302 (2004)}

\bibitem{zurek1984reversibility}
W. H. Zurek, 
{\it Reversibility and stability of information processing systems},
\href{https://journals.aps.org/prl/abstract/10.1103/PhysRevLett.53.391}{Phys. Rev. Lett. {\bf 53}, 391 (1984)}

\bibitem{zbarenco1997stabilization}
A. Barenco, A. Berthiaume, D. Deutsch, A. Ekert, R. Jozsa and C.  Macchiavello,
{\it Stabilization of quantum computations by symmetrization},
\href{https://epubs.siam.org/doi/10.1137/S0097539796302452}{SIAM J. Comput. {\bf 26},  1541 (1997)}

\bibitem{vaidman1996error}
L. Vaidman, L. Goldenberg and S. Wiesner,
{\it Error prevention scheme with four particles},
\href{https://journals.aps.org/pra/abstract/10.1103/PhysRevA.54.R1745}{Phys. Rev. A {\bf 54},  R1745 (1996)}

\bibitem{pachos2002quantum}
J. Pachos and H. Walther,
{\it Quantum computation with trapped ions in an optical cavity},
\href{https://journals.aps.org/prl/abstract/10.1103/PhysRevLett.89.187903}{Phys. Rev. Lett. {\bf 89}, APS, 187903 (2002)}

\bibitem{ralph2003quantum}
T. C. Ralph, A. Gilchrist, G. J. Milburn, W. J. Munro and S. Glancy,
{\it Quantum computation with optical coherent states},
\href{https://journals.aps.org/pra/abstract/10.1103/PhysRevA.68.042319}{Phys. Rev. A {\bf 68}, 042319 (2003)}

\bibitem{viola1998}
L. Viola and S. Lloyd,
{\it Dynamical suppression of decoherence in two-state quantum systems},
\href{https://journals.aps.org/pra/abstract/10.1103/PhysRevA.58.2733}{Phys. Rev. A {\bf 58}, 2733 (1998)}

\bibitem{vitali1999}
D. Vitali and P. Tombesi,
{\it Using parity kicks for decoherence control},
\href{https://journals.aps.org/pra/abstract/10.1103/PhysRevA.59.4178}{Phys. Rev. A {\bf 59},  4178 (1999)}

\bibitem{viola1999}
L. Viola, S. Lloyd  and  E. Knill,
{\it Universal control of decoupled quantum systems},
\href{https://journals.aps.org/prl/abstract/10.1103/PhysRevLett.83.4888}{Phys. Rev. Lett. {\bf 83},  4888 (1999)}

\bibitem{protopopescu2003}
V. Protopopescu, R. Perez, C. D'Helon  and J. Schmulen,
{\it Robust control of decoherence in realistic one-qubit quantum gates},
\href{https://iopscience.iop.org/article/10.1088/0305-4470/36/8/314}{J. Phys. A {\bf 36},  2175 (2003)}

\bibitem{bharti2022}
K. Bharti et al.
{\it Noisy intermediate-scale quantum algorithms}, 
\href{https://link.aps.org/doi/10.1103/RevModPhys.94.015004}{Rev. Mod. Phys. {\bf 94}, 015004 (2022)}.

\bibitem{barik2020}
S. Barik, D. K. Kalita, B. K. Behera, P. K. Panigrahi,
{\it Demonstrating quantum zeno effect on IBM Quantum Experience},
\href{https://arxiv.org/abs/2008.01070}{arXiv preprint arXiv:2008.01070 (2020)}.

\bibitem{thesis}
J. Sudhanva,
{\it Realization of Quantum Zeno Effect on IBM Quantum Experience and QISKIT},
\href{https://www.researchgate.net/publication/363639456_Realization_of_Quantum_Zeno_Effect_on_IBM_QUANTUM_Experience_and_QISKIT}{DOI:10.13140/RG.2.2.26141.18409}.

\bibitem{dominik2018}
D. \ifmmode \check{S}\else \v{S}\fi{}afr\'anek, S. Deffner,
{\it Quantum Zeno effect in correlated qubits},
\href{https://link.aps.org/doi/10.1103/PhysRevA.98.032308}{Phys. Rev. A {\bf 98}, 032308 (2018)}.

\bibitem{matsuzaki2010}
Y. Matsuzaki, S. Saito, K. Kakuyanagi, and K. Semba,
{\it Quantum Zeno effect with a superconducting qubit},
\href{https://link.aps.org/doi/10.1103/PhysRevB.82.180518}{Phys. Rev. B  {\bf 82},  180518 (2010)}.

\bibitem{kakuyanagi2015}
K. Kakuyanagi, T. Baba, Y. Matsuzaki,  H. Nakano, S. Saito  and K. Semba,
{\it Observation of quantum Zeno effect in a superconducting flux qubit},
\href{https://iopscience.iop.org/article/10.1088/1367-2630/17/6/063035}{New J. Phys. {\bf 17}, 063035 (2015)}.

\bibitem{harrington2017}
P. M. Harrington, J. T. Monroe and K. W. Murch, 
{\it Quantum Zeno Effects from Measurement Controlled Qubit-Bath Interactions},
\href{https://link.aps.org/doi/10.1103/PhysRevLett.118.240401}{Phys. Rev. Lett. {\bf 118},  240401 (2017)}.

\bibitem{devoret2004}
M. H. Devoret, A. Wallraff, J. M. Martinis,
{\it Superconducting qubits: A short review},
\href{https://arxiv.org/abs/cond-mat/0411174}{arXiv preprint cond-mat/0411174 (2004)}.

\bibitem{kockum2019}
A. F. Kockum, F. Nori,
{\it Quantum Bits with Josephson Junctions},
\href{https://doi.org/10.1007/978-3-030-20726-7_17}{Springer International Publishing, 703--741 (2019)}.

\bibitem{koch2007}
J. Koch, T. M. Yu, J. Gambetta, A. A. Houck, D. I. Schuster, J. Majer, A. Blais, M. H. Devoret, S. M. Girvin, R. J. Schoelkopf,
{\it Charge-insensitive qubit design derived from the Cooper pair box},
\href{https://link.aps.org/doi/10.1103/PhysRevA.76.042319}{Phys. Rev. A {\bf 76(4)}, 042319 (2007)}.

\bibitem{transpile}
IBM Quantum Documentation,
{\it Transpiler},
\href{https://docs.quantum.ibm.com/api/qiskit/transpiler}{https://docs.quantum.ibm.com/api/qiskit/transpiler}

\bibitem{swap}
Pennylane,
{\it What is a SWAP gate?},
\href{https://pennylane.ai/qml/glossary/what-is-a-swap-gate/}{https://pennylane.ai/qml/glossary/what-is-a-swap-gate/}

\bibitem{BCS}
Benenti, Giuliano and Casati, Giulio and Strini, Giuliano,
{\it Principles of Quantum Computation and Information - Volume II : Basic Tools and Special Topics},
\href{https://books.google.it/books?id=Its7DQAAQBAJ}{https://books.google.it/books?id=Its7DQAAQBAJ}

\bibitem{fakeprov}
IBM Quantum Documentation,
{\it Fake Provider},
\href{https://docs.quantum.ibm.com/api/qiskit/0.40/providers_fake_provider}{https://docs.quantum.ibm.com/api/qiskit/0.40/providers\_fake\_provider}

\bibitem{backendV2}
 IBM Quantum Documentation,
{\it BackendV2},
\href{https://docs.quantum.ibm.com/api/qiskit/0.40/qiskit.providers.BackendV2}{https://docs.quantum.ibm.com/api/qiskit/0.40/qiskit.providers.BackendV2}

\bibitem{calibrmatr}
IBM Quantum Documentation,
{\it qiskit.ignis.mitigation.CompleteMeasFitter},
\href{https://docs.quantum.ibm.com/api/qiskit/0.28/qiskit.ignis.mitigation.CompleteMeasFitter}{https://docs.quantum.ibm.com/api/qiskit/0.28/qiskit.ignis.mitigation.CompleteMeasFitter}

\bibitem{mitigation}
Yu, Hongye and Wei, Tzu-Chieh,
{\it Efficient separate quantification of state preparation errors and measurement errors on quantum computers and their mitigation}, 
\href{https://arxiv.org/abs/2310.18881}{arXiv preprint arXiv:2310.18881 (2023)}.

\bibitem{ECR}
IBM Quantum Documentation,
{\it ECRGate}, 
\href{https://docs.quantum.ibm.com/api/qiskit/qiskit.circuit.library.ECRGate}{https://docs.quantum.ibm.com/api/qiskit/qiskit.circuit.library.ECRGate}







\end{thebibliography}
\end{document}